\documentclass[eqsecnum,preprint,prd,aps,referee]{revtex4}
\usepackage{amsmath}
\usepackage{graphics}
\usepackage{epsfig}
\newcommand{\be}{\begin{equation}}
\newcommand{\ee}{\end{equation}}
\newcommand{\ba}{\begin{eqnarray}}
\newcommand{\ea}{\end{eqnarray}}
\newcommand{\bc}{\begin{center}}
\newcommand{\ec}{\end{center}}

\def\be{\begin{equation}}
\def\ee{\end{equation}}

\begin{document}
\begin{center}
\bibliographystyle{article}


{\Large \textsc{Noether symmetry for Gauss-Bonnet dilatonic gravity }}
\end{center}


\date{\today}

\author{Abhik Kumar Sanyal,$^{1}$  Claudio Rubano$^{3}$   and Ester Piedipalumbo$^{2,3}$
 \footnote{electronic-mail:\\
 author for correspondence: ester@na.infn.it\\ sanyal\_ak@yahoo.com,\\
rubano@na.infn.it}}
\affiliation{${\ }^{1}$ Dept. of Physics, Jangipur College, Murshidabad,  West
Bengal, India - 742213\\
${\ }^{2}$Dipartimento di Scienze Fisiche, Complesso Universitario di Monte
S. Angelo,\\
Via Cintia, Edificio N, 80126 Napoli, Italy\\
${\ }^{3}$Istituto Nazionale di Fisica Nucleare, Sezione di Napoli,\\
Complesso Universitario di Monte S. Angelo, Via Cintia, Edificio N, 80126
Napoli, Italy}

\begin{abstract}
Noether symmetry for Gauss-Bonnet-Dilatonic interaction exists for a
constant dilatonic scalar potential and a linear functional dependence of the
coupling parameter on the scalar field. The symmetry with the same form of the
potential and coupling parameter exists all in the vacuum, radiation and
matter dominated era. The late time acceleration is driven by the effective
cosmological constant rather than the Gauss-Bonnet term, while the later
compensates for the large value of the effective cosmological constant giving
a plausible answer to the well-known coincidence problem.
\end{abstract}
\maketitle

PACS number(s): 98.80.Jk, 98.80.Cq, 98.80.Hw, 04.20.Jb

KEYWORDS: Theoretical cosmology, Dark Energy, Observational Cosmology

\section{Introduction}

Many alternative theories of gravity have been proposed so far in
order to present viable cosmological models of dark energy
associated with observed cosmic acceleration. Among them, the
introduction of a Gauss-Bonnet term into the Gravitational action
has received much attention in recent years
\cite{a1}\cite{a2}\cite{a3}\cite{a4}\cite{a5}\cite{a6}\cite{a7}\cite{a8}
\cite{a9}\cite{a10}\cite{a11}\cite{a12}\cite{a13}\cite{a14}\cite{a15}\cite{od1},\cite{od2}.
In particular, important issues like - late time dominance of dark
energy after a scaling matter era and thus alleviating the
coincidence problem, crossing the phantom divide line and
compatibility with the observed spectrum of cosmic background
radiation have also been addressed recently \cite{b1}\cite{b2}.
Gauss-Bonnet term arises naturally as the leading order of the
$\alpha ^{\prime}$ expansion of heterotic superstring theory,
where, $\alpha^{\prime}$ is the inverse string tension.
\cite{c1}\cite{c2}\cite{c3}\cite{c4}.\newline However,
Gauss-Bonnet is a topologically invariant term in 4-d space-time
and so it is coupled with a dilatonic scalar field\textbf{,} in
order to avoid collapse of the equations to those corresponding to
standard cosmological model. As a result, at least two unknown
functions are to be postulated, or derived, viz., the potential of
the scalar field $V(\phi)$ and the coupling of the Gauss-Bonnet
term with gravity $\Lambda(\phi)$. A most elegant procedure is to
make a single postulate in order to derive these two functions
rather than setting both of them arbitrarily by hand. This may be
done by demanding Noether symmetry amongst field variables, which
has received much attention in recent times, particularly in the
context of higher order theory of gravity
\cite{d1}\cite{d2}\cite{d3}.\newline\noindent The Noether symmetry
approach for the solution of the cosmological equations was
developed many years ago \cite{e1}, and since then applied to find
general exact solutions of many problems in the field
\cite{e2}\cite{e3}\cite{e4}\cite{e5}\cite{e6}\cite{d1}. It
consists first in recognizing that the field equations, when turn
out to be ordinary differential equations, may be derived from an
ordinary point Lagrangian. Then, it is required to select the (not
yet established) functions under the condition that the Lagrangian
should be preserved under some infinitesimal point transformation
(Lie derivative). Once the functions are obtained, general exact
integration of the field equations may be usually performed. If
not, it simplifies the set of differential equations considerably,
as in the present case, which helps in discussing the
solutions and setting the values of parameters of the theory. The
discussion on the physical implications of this symmetry may be
found in  \cite{e2}. As a matter of fact, its nature remains
obscure, but it revealed so fruitful in may circumstances  that it is worth to attempt its applicability hereto. Of course, here like earlier works, the same results may be obtained by suitable guess of the functions and transformations. However it should be made clear that
it is extremely difficult to make such a guess without Noether symmetry approach.

In the present work, our starting point is the gravitational
action with Gauss-Bonnet term being coupled with a dilatonic
scalar, in the presence of cold dark matter, for which Noether
symmetry has been explored in the background of spatially flat
Robertson-Walker metric. Noether symmetry has been found in the
matter dominated era, after setting the state parameter $w$
corresponding to the baryonic and the cold dark matter to zero. In
the process the potential has been fixed to a constant and the Gauss-Bonnet
coupling parameter $\Lambda(\phi)$ turned out to be a linear
function of $\phi$. In the subsection 2.1, we have generated a set
of solutions simply by handling the algebraic equation in Hubble
parameter rather than solving differential field equations. The results
thus obtained are intriguing.\newline The original idea to include
Gauss-Bonnet term into the action was to drive the late time
cosmic acceleration, playing thus the key role of effective "dark
energy", instead of the scalar field. On the contrary, we
discovered that, at least in the circumstances described below, it
is again the scalar field which drives the acceleration. The
Gauss-Bonnet term instead plays the role of contrasting the
effective cosmological constant, ie., nullifying the effective
cosmological constant as should be clear below. \ However, to
reduce the cosmological constant by some $120$ order magnitude
requires $\Omega_{\phi}$ of the same order of magnitude. Though
there has been some early attempts in this regard \cite{r1,r2},
nevertheless it is an interesting issue, since, \ it has been
observed that modified gravity theory can contrast the effective
cosmological constant. In section 3 we compare our model with
$\Lambda$CDM, which as usual, shows that it is practically
impossible to identify the two, as far as luminosity distance
versus redshift graph is concerned. It is worth noting that the Noether symmetry approach could be applied also in the case of modified GB gravity theories, which includes, for instance, the functional dependence from the Gauss-Bonnet invariant G in the form of f(G) only, or also an additional dependence on curvature as f(G,R). Moreover, since Gauss-Bonnet is not a topologically invariant term in dimensions greater than 4, so we could also consider the standard Gauss-Bonnet gravity with or without dilatonic coupling (\cite{od3}). But then, it should be clear, however, that the mathematical feasibility of the method will be different and of course very difficult.

\section{The Model with Gauss-Bonnet Interaction and Noether symmetry}

We start with the following action containing Gauss-Bonnet interaction%

\begin{equation}
S=\int d^{4}x\sqrt{-g}[\frac{R}{2\kappa^{2}}+\frac{\Lambda(\phi)}{8}%
G(R)-\frac{1}{2}\phi_{;\mu}\phi^{;}{}^{\mu}-V(\phi)+L_{m}],
\end{equation}
where,
\[
G(R)=R^{2}-4R_{\mu\nu}R^{\mu\nu}+R_{\mu\nu\rho\sigma}R^{\mu\nu\rho\sigma}%
\]

\noindent is the Gauss-Bonnet term, which appears in the action with a
coupling parameter $\Lambda(\phi)$, $L_{m}$ is the matter Lagrangian and
$V(\phi)$ is the dilatonic potential. For the spatially flat Robertson-Walker
space-time $(k=0)$,
\[
ds^{2}=-dt^{2}+a^{2}(t)[dr^{2}+r^{2}d\theta^{2}+r^{2}sin^{2}\theta d\phi
^{2}],
\]
the field equations in terms of the Hubble parameter $H=\frac{\dot{a}}{a}$, are%

\begin{eqnarray}
&& 2\dot{H}+3H^{2}=-\left[  \frac{1}{2}\dot{\phi}^{2}-V(\phi)+2\Lambda^{\prime
}\dot{\phi}(H\dot{H}+H^{3})+(\Lambda^{\prime}\ddot{\phi}+\Lambda^{\prime
\prime}\dot{\phi}^{2})H^{2}+p_{m}\right]  =\nonumber\\
&& -(p_{GB}+p_{\varphi}+p_{m}),
\end{eqnarray}

\begin{equation}
3H^{2}=\left[  \frac{1}{2}\dot{\phi}^{2}+V(\phi)-3\Lambda^{\prime}\dot{\phi
}H^{3}+\rho_{m}\right]  =(\rho_{GB}+\rho_{\varphi}+\rho_{m}),\label{hubble}%
\end{equation}

\begin{equation}
\ddot\phi+3H\dot\phi+V^{\prime}=3\Lambda^{\prime}H^{2}(\dot H+H^{2}),
\end{equation}

\noindent in the units $\kappa^{2}(=8\pi G)=c=1$. Thus, $p_{GB}=$
$2\Lambda^{\prime}\dot{\phi}(H\dot{H}+H^{3})+(\Lambda^{\prime}\ddot{\phi
}+\Lambda^{\prime\prime}\dot{\phi}^{2})H^{2}$ and $\rho_{GB}=-3\Lambda
^{\prime}\dot{\phi}H^{3}$ are the effective pressure and the energy density
generated by the Gauss-Bonnet-scalar interaction, $p_{\varphi}=$ $\frac{1}%
{2}\dot{\phi}^{2}-V(\phi)$ and $\rho_{\varphi}=\frac{1}{2}\dot{\phi}%
^{2}+V(\phi)$, while $p_{m}$ and $\rho_{m}$ are the pressure and
the energy density , corresponding to background matter
distribution respectively. Equation (\ref{hubble}) provides the
standard constraint for the $\Omega$ parameters:
$1=\Omega_{\phi}+\Omega_{GB}+\Omega_{m}.$

Let us note that while in the standard quintessence scenario, where the dark
energy is described by means of a time decaying scalar field, as well in the
case of a bare cosmological constant, the nowadays energy density $\rho_{\phi
}$ is of the order of the present value of the Hubble parameter, so that
$\Omega_{\phi_{0}}\simeq 0.7$. Here in the presence of the Gauss-Bonnet
interaction, $\Omega_{GB_{0}}$ and $\Omega_{\phi_{0}}$ can contrarywise be
large, but it turns out that $\Omega_{GB_{0}}+\Omega_{\phi_{0}}\sim0.7$, as we
will explicitly investigate in subsection \ref{solutions}, and it is also
shown in Figure (2). The background matter satisfies the conservation law and
state equation%

\begin{equation}
\rho_{m}=Ma^{-3(1+w)}\quad,\quad p_{m}=w\rho_{m}%
\end{equation}

\noindent respectively, where  $M$ is a constant and $w$ is the
state parameter of the background matter. As usual in Noether
symmetry approach, we derive the field
equations from a point Lagrangian, which may be expressed as%

\begin{equation}
L=-3a\dot{a}^{2}-\Lambda^{\prime}\dot{\phi}\dot{a}^{3}+\frac{1}{2}a^{3}%
\dot{\phi}^{2}-a^{3}V(\phi)-Ma^{-3w}.
\end{equation}

Now, we apply the Noether symmetry approach. Let $\ \tilde{X}$ \ be a vector
field on the configuration space (minisuperspace) $\{a,\varphi\}$ of the
dynamical system%
\begin{equation}
\tilde{X}=\alpha(a,\varphi)\frac{\partial}{\partial a}+\beta(a,\varphi
)\frac{\partial}{\partial\varphi},
\end{equation}
where $\alpha,\beta$ are to be determined. $\tilde{X}$ can be lifted to the
tangent space $\{a,\varphi,\dot{a},\dot{\varphi}\}$ in the following standard
manner%
\begin{equation}
X=\alpha(a,\varphi)\frac{\partial}{\partial a}+\beta(a,\varphi)\frac{\partial
}{\partial\varphi}+\dot{\alpha}(a,\varphi)\frac{\partial}{\partial\dot{a}%
}+\dot{\beta}(a,\varphi)\frac{\partial}{\partial\dot{\varphi}},
\end{equation}
where
\begin{equation}
\dot{\alpha}=\frac{\partial\alpha}{\partial a}\dot{a}+\frac{\partial\alpha
}{\partial\varphi}\dot{\varphi}\quad;\quad\dot{\beta}=\frac{\partial\beta
}{\partial a}\dot{a}+\frac{\partial\beta}{\partial\varphi}\dot{\varphi}.
\end{equation}

Let us now demand Noether symmetry by imposing the condition%

\begin{equation}
\pounds _{X}L=XL=\alpha\frac{\partial L}{\partial a}+\beta\frac{\partial
L}{\partial\phi}+\dot\alpha\frac{\partial L}{\partial\dot a }+\dot\beta
\frac{\partial L}{\partial\dot\phi} = 0,
\end{equation}

\noindent where $\pounds _{X}L$ is the Lie derivative of the point Lagrangian
w.r.t. $X$.  Thus we have%

\begin{eqnarray}
&& \alpha\left(  -3\dot{a}^{2}+\frac{3}{2}a^{2}\dot{\phi}^{2}-3a^{2}
V+3wMa^{-3w-1}\right)  +\beta\left(  -\Lambda^{\prime\prime}\dot{\phi}
\;\dot{a}^{3}-a^{3}V^{\prime}\right)  +\\
&& +\left(  \frac{\partial\alpha}{\partial a}\dot{a}+\frac{\partial\alpha
}{\partial\phi}\dot{\phi}\right)  \left(  -6a\dot{a}-3\Lambda^{\prime}%
\dot{\phi}\dot{a}^{2}\right)  +\left(  \frac{\partial\beta}{\partial a}\dot
{a}+\frac{\partial\beta}{\partial\phi}\dot{\phi}\right)  \left(
-\Lambda^{\prime}\dot{a}^{3}+a^{3}\dot{\phi}\right)  =0.\nonumber
\end{eqnarray}

\noindent This equation  is satisfied provided the co-efficients of $\dot
{a}^{2}$, $\dot{\phi}^{2}$, $\dot{\phi}\dot{a}$, $\dot{\phi}\dot{a}^{3}$,
$\dot{\phi}^{2}\dot{a}^{2}$, $\dot{a}^{4}$ and the terms free from time
derivative vanish separately, ie.,

\begin{equation}
-3\alpha-6a\frac{\partial\alpha}{\partial a}=0\Longrightarrow\alpha
+2a\frac{\partial\alpha}{\partial a}=0,\label{N11}
\end{equation}

\begin{equation}
\frac{3}{2}a^{2}\alpha+a^{3}\frac{\partial\beta}{\partial\phi}%
=0\Longrightarrow3\alpha+2a\frac{\partial\beta}{\partial\phi}=0,\label{N12}%
\end{equation}

\begin{equation}
-6a\frac{\partial\alpha}{\partial\phi}+a^{3}\frac{\partial\beta}{\partial
a}=0\Longrightarrow6\frac{\partial\alpha}{\partial\phi}-a^{2}\frac
{\partial\beta}{\partial a}=0\label{N13}
\end{equation}

\begin{equation}
\Lambda^{\prime\prime}\beta+\Lambda^{\prime}(3\frac{\partial\alpha}{\partial
a}+\frac{\partial\beta}{\partial\phi})=0,\label{N14}%
\end{equation}

\begin{equation}
-3\Lambda^{\prime}\frac{\partial\alpha}{\partial\phi}=0,\label{N15}
\end{equation}

\begin{equation}
-\Lambda^{\prime}\frac{\partial\beta}{\partial a}=0,\label{N16}
\end{equation}

\begin{equation}
3\alpha\lbrack wMa^{-3w-1}-a^{2}V]-a^{3}V^{\prime}\beta=0\left(
\Longrightarrow\beta\frac{V^{\prime}}{V}=-3\frac{\alpha}{a},\;{\rm
for}\;\;w=0\right) .\label{N17}
\end{equation}

\noindent It was already pointed out in the introduction that if $\Lambda
(\phi)=0$, field equations collapses to standard cosmological ones, so that
one recovers already known results; thus $\Lambda\neq0$. As a result,
equations (\ref{N15}) and (\ref{N16}) immediately reveal that $\alpha
=\alpha(a)$ and $\beta=\beta(\phi)$, which satisfy equation (\ref{N13}).
Hence, equation (\ref{N11}) implies, $\alpha=K/\sqrt{a}$, where, $K$ is a
constant. However, equation (\ref{N12}) is satisfied only for $K=0$. We thus have,%

\begin{equation}
\alpha= 0,
\end{equation}

\noindent and so, equation (\ref{N17}) implies, $\beta\frac{V^{\prime}}{V}=0$.
The vector field $X$ vanishes for $\beta=0$, and symmetry remains obscure, so
we must have,

\begin{equation}
V=constant=V_{0}~.
\end{equation}

\noindent Thus we obtain  a rather strong result viz., the potential is obliged
to be a constant. So the scalar field energy density  $\rho_{\phi}$ becomes
more properly an effective cosmological constant $\Lambda_{eff(\varphi)}%
=\frac{1}{2}\dot{\phi}^{2}+V_{o}$. If $\dot{\phi}$ falls off with world time,
one is left with the bare cosmological constant.  In Fig.(1) we show the
evolution of such term for some suitable  values of parameters (chosen in Sec.
2.1). The total, i.e. the  observed  effective cosmological constant in the
present context is   $\Lambda_{eff}=\Lambda_{eff(\varphi)}+\Lambda$.

However, even more interesting result that we observe at this stage is that,
$\alpha$ has to vanish for the existence of Noether symmetry. Thus, if one
considers the term in brackets of equation (\ref{N17}), it is clear that the
matter state parameter $w$ is irrelevant, as well as a possible non zero value
of the space curvature $k$,  so that $V=V_{0}$ remains constant for the whole
evolutionary history of the Universe, ie., in the vacuum and the radiation
dominated era. Thus, the following charge remains conserved also. In any case,
in the present work we deal with late time evolution of the Universe and will
not insist on early Universe models.
\begin{figure}[ptb]
\begin{center}
\includegraphics[
height=1.9104in, width=2.968in] {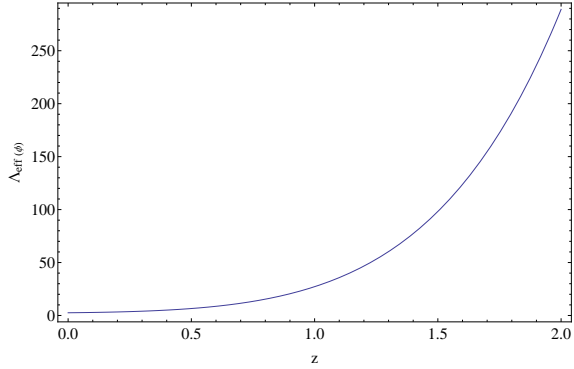}
\end{center}
\caption{Behaviour of the $\Lambda_{eff}$ term with the redshift (the
parameters entering in the model are $\Omega_{m}=0.33$,\thinspace\ $H_{0}%
=1$,\thinspace\ $\lambda=0.2$ }%
\label{leff}%
\end{figure}\noindent Now in view of equation (\ref{N12}), we get a constant
value for $\beta=\beta_{0}$.\noindent\ Equation (\ref{N14}) then yields
$\Lambda^{\prime\prime}=0$, ie.,%

\begin{equation}
\Lambda= \lambda\phi,
\end{equation}

\noindent with constant $\lambda$. Thus, in view of Noether
symmetry, we have been able to find the functional forms of
$\Lambda(\phi)$ and $V(\phi)$ of the model under consideration. It is worth noting that one may always choose $V = V_0 = \rm{constant}$, a-priori, but choosing a linear form of $\Lambda(\phi)$, without invoking Noether symmetry is highly ambitious. Further, such choices are usually made for mathematical simplicity. We find here that, although apparently simple forms have emerged from Noether symmetry,
 however, it do not make the field equations tractable to solve directly. It is again interesting to note that, the same form of $\Lambda(\phi)$ was obtained in an earlier work \cite{b2} at the
late time cosmic evolution, when the kinetic term viz.,
$\dot{\phi}^{2}$ indeed became constant and the Universe sets off
for an accelerated expansion. We just like to mention once again
that the coupling parameter and the potential thus found are
independent of the evolutionary history of the Universe.

\noindent Finally, the point Lagrangian takes the following form,%

\begin{equation}
L = -3 a\dot a^{2}-\lambda\dot\phi\dot a^{3}+\frac{1}{2}a^{3}\dot\phi
^{2}-a^{3} V_{0} - M,
\end{equation}

\noindent which is clearly cyclic in $\phi$. We do not need thus to perform a
transformation of variables, as is usually done in Noether symmetry approach
\cite{e1}. The associated conserved quantity is found quite trivially as%

\[
n=a^{3}\dot{\phi}-\lambda\dot{a}^{3},
\]

\noindent which is essentially the generalization of a well known result in
the case of constant potential \cite{e1}. The field equations now are the following,

\begin{equation}
2\dot{H}+3H^{2}=-\left(  \frac{1}{2}\dot{\phi}^{2}-V_{0}+2\lambda\dot{\phi
}(H\dot{H}+H^{3})+(\lambda\ddot{\phi})H^{2}\right)  ,\label{N22}%
\end{equation}

\begin{equation}
3H^{2}=\frac{1}{2}\dot{\phi}^{2}+V_{0}-3\lambda\dot{\phi}H^{3}+\frac{M}{a^{3}%
},\label{N23}%
\end{equation}

\begin{equation}
a^{3}\dot{\phi}-\lambda\dot{a}^{3}=n,\label{N24}%
\end{equation}

\noindent\ However, only two of these are independent and one can utilize the
last two for finding explicit solutions of the scale factor and the scalar
field, which will finally set the coupling parameter $\Lambda(\phi)$. For this
purpose, let us substitute $\dot{\phi}$ from equation (\ref{N24}) in equation
(\ref{N23}), to get,

\begin{equation}
3\frac{\dot{a}^{2}}{a^{2}}=-\frac{5}{2}\lambda^{2}\frac{\dot{a}^{6}}{a^{6}%
}-2\lambda n\frac{\dot{a}^{3}}{a^{6}}+\frac{n^{2}}{2a^{6}}+\frac{M}{a^{3}%
}+V_{0},\label{N25}%
\end{equation}

\noindent which may be rewritten as,

\begin{equation}
\frac{5}{2}\lambda^{2}\dot{a}^{6}+2\lambda n\dot{a}^{3}+3a^{4}\dot{a}%
^{2}-V_{0}a^{6}-Ma^{3}-\frac{n^{2}}{2}=0.\label{N26}%
\end{equation}

\noindent This equation is definitely not easy to solve, so in the following
section, we take up a different route to analyze the solutions.

\subsection{Obtaining solutions}

\label{solutions}

As mentioned, it is clearly not possible to solve analytically equation
(\ref{N26}) in order to obtain $a(t)$, but, on the other hand, it turns out to
be unnecessary. It can be transformed into an algebraic equation for the
Hubble parameter, being interpreted as a function of the red-shift $z$, which
is clearly what we need in order to investigate the cosmic evolution in the
context of dark energy. Thus, we get,

\begin{equation}
5\lambda^{2}H^{6}+4n\lambda(1+z)^{3}H^{3}+6H^{2}-n^{2}(1+z)^{6}-2M(1+z)^{3}%
-2V_{0}=0,\label{N27}%
\end{equation}

\noindent where we have set the present value of the scale factor $a_{0}=1$,
as usual. The only problem left is the impossibility to get an exact solution
of a 6th degree algebraic equation. We must thus give effort to make a
reasonable choice of the parameters involved, in order to make a somewhat detailed study of the cosmological consequence of the situation under investigation. Let us stress that, in any case, we are not talking of numerical
integration but of simple solution of an algebraic equation.

\noindent First, since we want to investigate on acceleration at the present
epoch, we need an expression for $H^{\prime}(z)$ (from now on, prime means
derivative w.r.t. $z$). Taking derivative of equation (\ref{N27}) it is easy
to obtain,

\begin{equation}
H^{\prime}(z)=\frac{6n^{2}(1+z)^{5}+6M(1+z)^{2}-12n\lambda(1+z)^{2}H^{3}%
}{30\lambda^{2}H^{5}+12n\lambda(1+z)^{3}H^{2}+12H}.\label{N28}
\end{equation}

\noindent Second, our choice of units leave us free to choose the unit of
time. In a recent work, some of us \cite{f} fixed the present age of the
universe $t_{0}=1$. This was due to the fact that, in that case, we had
explicit time dependance of the solutions. Here, instead we  set the Hubble
time to one, i.e., $H_{0}=1$. It  should be clear that this does not imply any
loss of generality.

\noindent Third, we need some ``reasonable values'' for other
parameters. By this we mean to set some parameters of the theory
in such a way as to get simple expressions for computations,
ending up with a model reasonably similar to the present
observable universe, although may not be the best fit. This is due
to the fact that for the moment we are mostly interested here to
show some important features implied by the introduction of a GB
term into the action. A more precise statistical treatment will be
given in sec. \ref{snIa}.

\noindent We observe that $M$ parameterizes the amount of matter. In our units
it is simply $M=3\Omega_{m0}$. Thus if we assume $\Omega_{m0}=1/3$, we obtain
a nice value $M=1$. Since  our model is different from $\Lambda$CDM, it is by
far not sure that we should obtain the current value of that model, i.e.
$\Omega_{m0}=0.26$. This sort of arbitrariness will be fixed later.

\noindent The second reasonable choice is to assume that the present value of
the deceleration parameter is $q_{0}=-1/2$ (again with some arbitrariness).
Hence, finally we are left with

\begin{equation}
V_{0}=2\mp\frac{1}{4}\left(  \sqrt{19}\mp8\right)  \lambda^{2}>0\quad\,;\quad
n=\mp\frac{1}{2}\left(  \sqrt{19}\mp3\right)  \lambda.\label{V0n}%
\end{equation}

\noindent We observe that we are finally left with only the $\lambda$
parameter, that will eventually fix up $V_{o}$ and $n$, but with two possible choices, corresponding to two different signs. In the following we will adopt the choice corresponding to the plus signs; however the  other one, corresponding to minus signs into equation (\ref{V0n}) turns out to be also interesting. Another interesting
remark is that $V_{0}$ cannot be zero, attaining a minimum value of 2, for
$\lambda=0$. We have also  checked that, in general,  it is impossible to
obtain any acceleration with zero value for $V_{0}$. This means that,
actually, it is  an effective cosmological constant $\Lambda_{eff(\varphi)}$,
which drives the acceleration. Thus one may ask what is the point in setting
up all this stuff if the final answer is that we must stay again with the old
good $\Lambda$ ? For a possible answer now let us eventually go to some
physical quantities.

\noindent According to the present choice,  we have already fixed $\Omega_{m_0} =\frac{1}{3}$, and so are required to compute
$\Omega_{\phi0}$ and $\Omega_{GB0}$.  With our choice of parameters we get
\begin{equation}
\Omega_{GB}=-\frac{\lambda(\lambda H^{3}+n/a^{3})}{H^{3}}\quad\,;\quad
\Omega_{\varphi}=\frac{2}{3}-\Omega_{GB},\label{N29b}%
\end{equation}
and, substituting $n$ from equation (\ref{V0n}) and $H_{0}=a_{0}=1$, we obtain
\begin{equation}
\Omega_{GB0}=\frac{1}{2}\left(  \sqrt{19}-5\right)  \lambda^{2}\quad
\,;\quad\Omega_{\varphi0}=\frac{1}{6}\left(  4-3\left(  \sqrt{19}-5\right)
\lambda^{2}\right)  .\label{N29c}%
\end{equation}
Let us remind that $\Omega_{\phi0}=\rho_{\phi}/(3H_{0}^{2})$ may be
interpreted as an effective cosmological constant, so that its evaluation is
very important. As $\Omega_{\phi0}+\Omega_{GB0}=\frac{2}{3}$, there is clearly
a degeneration, between $\Omega_{\phi0}$ and $\Omega_{GB0}$. It is interesting
to look at a compared plot of the two.

\begin{figure}[ptb]
\label{omegasfig}
\par
\begin{center}
\includegraphics[
height=1.9104in, width=2.968in] {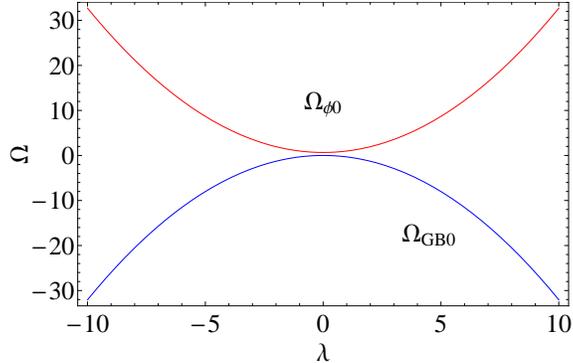}
\end{center}
\caption{$\Omega_{\varphi0}$ (upper, red) and $\Omega_{GB0}$(lower, blue)
versus $\lambda$. Please note that there is a  small gap between  the two curves. From text it should be clear that they never sum up to zero}%
\end{figure}

\noindent Figure (2) depicts that small values of $\lambda$ takes care of
small values of the "effective cosmological constant". Instead, large and
possibly huge values allow for large (huge) values of it. Thus we see that the
presence of the Gauss-Bonnet-dilatonic interaction term $\lambda$ gives us the
possibility to reduce the tremendous repulsive power of a large effective
cosmological constant $\Lambda_{eff(\varphi)}$. {Hence the well known
coincidence problem viz., why the cosmological constant is so small today
might have been given a plausible answer. This of course is true if all this
mechanism gives a good fit with data, which we do just in the next section.

\section{Comparison with $\Lambda CDM$ and with the observational SnIa data}

\label{snia}

We are now ready to compare our model with $\Lambda$CDM, to show that they are
observationally equivalent, as far as luminosity distance is under
consideration. The value of $\lambda$ should be  irrelevant in this context.
We have checked that it is indeed so, and present a result with $\lambda=0.2$.

\noindent Let us consider the standard $\Lambda$CDM expression of the Hubble
parameter, normalized to $H_{0}=1$, as mentioned above
\begin{equation}
H_{\Lambda CDM}=\sqrt{\Omega_{\Lambda CDM}(1+z)^{3}+1-\Omega_{\Lambda CDM}},
\end{equation}
and compare it with our model. The values for $H$ are obtained by
means of numerical solution of equation (\ref{N27}) point by
point, with some care in the treatment of the branching points.
The best we can do is for $\Omega_{\Lambda CDM}=0.7 .$

\begin{figure}[ptb]
\begin{center}
\includegraphics[
height=2.2278in, width=3.3797in] {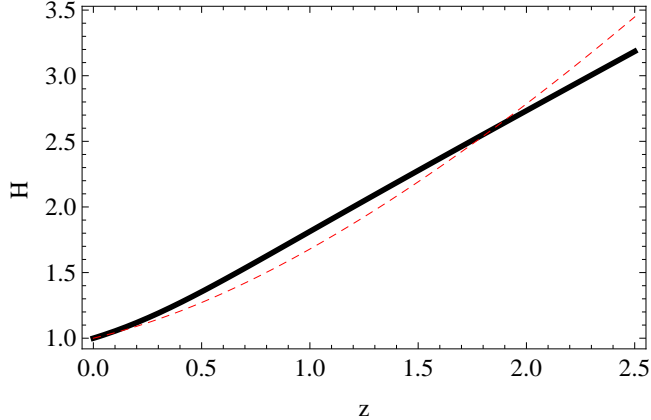}
\end{center}
\caption{$H_{\Lambda CDM}$ (dotted red line)compared with our model (continuous
line). The parameters entering in the model are $\Omega_{m}=0.33$,\thinspace\ $H_{0}%
=1$,\thinspace\ $\lambda=0.2$ }%
\end{figure}

\noindent Figure (3) shows that they look rather different. But we know that
\begin{equation}
D_{L}(z)=(1+z)\int_{0}^{z}{\frac{dz^{\prime}}{H(z^{\prime})}},\nonumber
\end{equation}
so that the passage to luminosity distance and then to distance modulus \textbf{act}
in killing the differences. In our case, the first step is enough. It is
possible to compute numerically both luminosity distances and obtain the
following plot in Figure (4) which shows that the overlap is perfect, and the
slight difference in the values of $\Omega$'s is irrelevant, as mentioned
above. \begin{figure}[ptb]
\begin{center}
\includegraphics[
height=2.1439in, width=3.3797in] {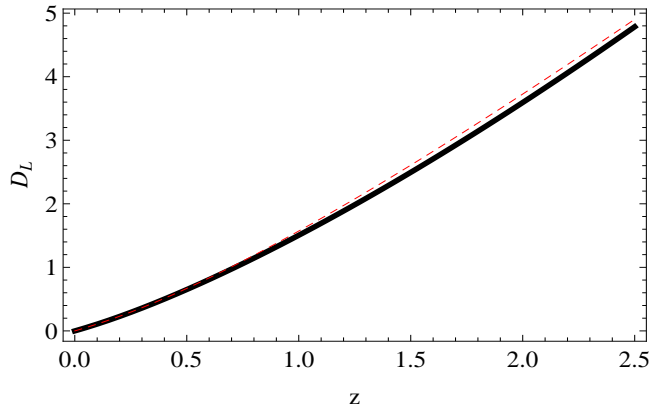}
\end{center}
\caption{Luminosity distances for $\Lambda$CDM (dotted line), with
$\Omega_{\Lambda}=0.73$, and our model (continuous line), with parameters entering  $\Omega_{m}=0.33$,\thinspace\ $H_{0}%
=1$,\thinspace\ $\lambda=0.2$ (corresponding to $\Omega_{\phi_{0}}+\Omega_{{GB}_0}=0.67$).}
\end{figure}

\subsection{Constraints from recent SNIa observations}

\label{snIa} In this subsection we show that our model is compatible with
recent observational data, in particular with the observations of type Ia
supernovae. We use the most updated SNeIa sample. The present compilation, which is referred to as Union \cite{Union2}, includes recent large samples from SNLS \cite{SNLS} and ESSENCE \cite{ESSENCE} surveys, older data sets and the recently extended data set of distant SNeIa, observed with HST. All of these have been homogeneously reanalyzed with the same lightcurve fitter. After
selection cuts and outliers removal, the final sample contains 557 SNeIa
spanning the range $0.015\leq z\leq1.55$. To constrain our models we actually
compare the \textit{theoretically \thinspace\ predicted} distance modulus
$\mu(z)$ with the \textit{observed}, through a likelihood analysis, where we
use as merit function the likelihood $\mathcal{L}=\exp{\left(  -\frac{1}
{2}\chi^{2}\right)  }$. The distance modulus is defined by
\begin{equation}
\mu=m-M=5\log{D_{L}(z)}+5\log({\frac{c}{100h}})+25,\label{eq:mMr}%
\end{equation}
where $m$ is the appropriately corrected apparent magnitude including
reddening, K correction etc.,  $M$ is the corresponding absolute magnitude,
$D_{L}$ is the luminosity distance in Mpc, and $h$ is the standard
dimensionless Hubble constant. We note that because of our choice of time
unit, our Hubble constant is not (numerically) the same as the $\tilde{H}_{0}$
that is usually measured in $\mathrm{kms^{-1}Mpc^{-1}}$. Actually, we may
easily obtain the relation
\begin{equation}
h=9.9{\frac{H_{0}}{\tau}}\,,\label{ha}%
\end{equation}
where as usual $h=\tilde{H}_{0}/100$ and $\tau$ is the age of the
universe in Gy, $\tilde{H}_{0.}$ being the Hubble parameter in
standard units. We see that $H_{0}$ fixes only the product
$h\tau$. In particular, we know that $\tau=13.73_{-0.15}^{+0.16}$
(see for instance \cite{WMAP3}), thus we get $h\leq 0.76$ for
$H_{0}\approx1$. Before going into the details of our statistical
analysis thoroughly, it is needed to turn  into the
parametrization of our solutions, mainly following the Eqs.
(\ref{V0n},\ref{N29b},\ref{N29c}). Actually, earlier (in the
previous subsection), in order to illustrate some basic properties
of our model, we fixed the values of $\Omega_{\mathrm{m0}}$, $q_0$
,$H_0$ and  $\lambda$ also. Here, since we want to constrain the
values of the physically meaningful parameters as maximum
likelihood \textbf{ones}, comparing theoretical predictions with
observational data, we remove the restrictions on the parameters.
More generally, indeed
\begin{eqnarray}\label{sum}
 && \Omega_{\phi_0}+\Omega_{{GB}_0}= \frac{1}{6} \left[-6 H_{0}^2 (\Omega_{m0}-1)+(H_{0}-1) (H_{0}+1) \left(H_{0}^2+H_{0}+1\right)\right.\nonumber \\&& \left.
  (H_{0} (9H_{0}-5)+5) \lambda ^2 -4 \left(H_0^3-1\right) H_{0}^2 q_0 \lambda ^2 + \right. \\&& \left. -4 \left(H_{0}^3-1\right) \lambda   \sqrt{H_{0}
   \left(H_{0}^3 \lambda ^2 \left(H_{0}^2-7 H_{0} (q_0 - 1)+(q_0-1)^2\right)-3 H_{0}
  \Omega_{m0}-2 q_0 +2\right)}\right].\nonumber
\end{eqnarray}
In fact  we have $ \Omega_{{\phi}_0}
+\Omega_{{GB}_0}=1-\Omega_{m0}$, as due, in order to satisfy the
Einstein equations. If $H_0\neq 1$, the previous relation has to
be considered as a constraint among the parameters. We actually
used it to  express $q_0$ as function of $H_0$, $\Omega_{m0}$ and
$\lambda$. The modulus of distance in Eq. (\ref{eq:mMr}) turns out
to be function of $z$, and $H_0$, $\Omega_{m0}$ and $\lambda$.
Performing our statistical analysis with the Union2 compilation we
marginalize over the $\lambda$ parameter, that is we maximize the
likelihood
$\mathcal{L}_{marg}=\int_{\lambda_{min}}^{\lambda_{max}}{d\lambda
\exp{\left(  -\frac{1}{2}\chi^{2}\right)  }}$, where
$\lambda_{min}$ and $\lambda_{max}$ are fixed by asking that $q_0$
should lie in the region allowed by the observations (see for
instance \cite{daly}). We obtain $\chi_{red}^2=0.99$ for 554
points, and the  best fit  values are $h=0.70_{-0.05}^{+0.03}$,
and $\Omega_{\mathrm{m0}}=0.35_{-0.08}^{+0.04}$. In Fig.
\ref{union_fit} we compare the best fit curve with the
observational dataset. Let us remark that the range of
$\Omega_{\mathrm{m0}}$ here obtained,
 includes the particular value
chosen in earlier section. Further, its lower limit is  consistent
with the presently acceptable value.  \begin{figure}[ptb]
\centering{  \includegraphics[ height=2.1439in,
width=3.5in]{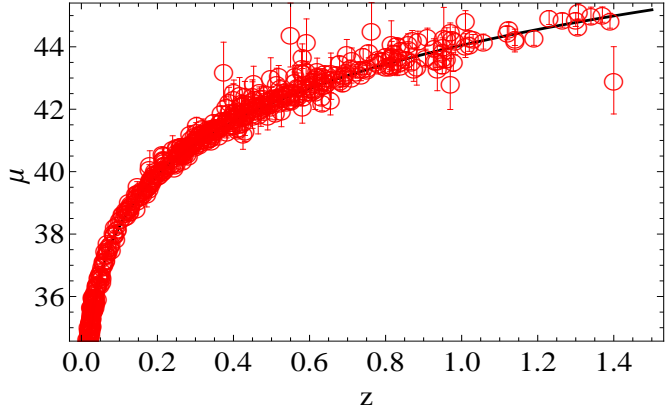}}  \caption{{\small Observational data
of the Union2 compilation fitted to our model. The red curve is
the best fit curve with $h=0.70_{-0.05}^{+0.03}$, and corresponds
to $\Omega
_{\mathrm{M0}}=0.35_{-0.08}^{+0.04}$ }}%
\label{union_fit}%
\end{figure}

\section{Conclusions}

Noether symmetry has been enforced in Gauss-Bonnet dilatonic scalar theory of
gravity and the following important results have emerged.

\noindent

\begin{description}
\item[ (1) ] The coupling parameter and the potential thus found
are independent of the evolutionary history of the Universe, ie.,
Noether symmetry exists throughout the history of evolution of the
Universe starting from the early vacuum dominated era, passing
over to radiation dominated era and finally at the matter
dominated era. Existence of such symmetry fixes up the
dilatonic-Gauss-Bonnet coupling parameter $\Lambda=\lambda\phi$
and the scalar potential $V=V_{0}$ (a constant) once and
forever.\newline

\item[ (2) ] The same form of $\Lambda(\phi)$ was obtained in an earlier work
\cite{b2} at the late time cosmic evolution, when the kinetic term viz.,
$\dot\phi^{2}$ indeed became constant and the Universe sets off for an
accelerated expansion.\newline

\item[ (3) ] Since the potential is obliged to be a constant, the effective
cosmological constant is now $\Lambda_{eff}=\Lambda_{eff(\varphi)}%
+\Lambda_{GB}$, which may be comparable with the present Hubble
parameter.\newline

\item[ (4) ] The late time cosmic acceleration is driven by the
scalar field effective cosmological constant rather than the
Gauss-Bonnet term.\newline

\item[ (5) ] Figure (2) depicts that the presence of the
Gauss-Bonnet-dilatonic interaction term $\lambda$ puts up the possibility to
reduce the tremendous repulsive power of a large effective cosmological
constant. Thus the  well known coincidence problem viz., why the cosmological
constant is so small today might have been given a plausible answer.\newline

\item[ (6) ] For late universe, there is an almost perfect equivalence with
the $\Lambda$CDM model as depicted in Figure (4).

\item[ (7) ] A more accurate fit procedure gives even more satisfactory
results, with a best fit value for $\Omega_{m}$, whose lower limit is
consistent with the results obtained from very different kinds of methods
using various astronomical objects, within the framework of the standard
$\Lambda CDM$ model \cite{WMAP3,sab}.
\end{description}

{\bf Acknowledgements:}

A.K. Sanyal is grateful to the University of Naples (Ufficio
Relazioni Internazionali) for supporting a visit in the Department
of Physical Sciences, Naples.

\end{document}